\documentclass[referee,traditabstract]{aa}
\usepackage{graphicx,ulem,amsmath,relsize,epsfig,epstopdf}
\usepackage{txfonts}
\usepackage{url}
\usepackage{booktabs}

\newcommand{\rs}{$R_{\odot}$}

\newcommand{\rmq}{rad m$^{-2}$}
\newcommand{\srmi}{$\sigma_{RM,i}$}
\newcommand{\srmo}{$\sigma_{i,obs}$}
\newcommand{\chitn}{$\chi^2_\nu$}
\newcommand{\chit}{$\chi^2$}
\newcommand{\omi}{$\omega_i$}
\newcommand{\drm}{$\delta RM$}
\newcommand{\ms}{Mancuso \& Spangler (2000)}

\begin{document}

\title{The radial profile of the inner heliospheric magnetic field as deduced \\ from Faraday rotation observations}
\author{S. Mancuso \inst{1} \and M. V. Garzelli \inst{2}}
\institute{INAF - Osservatorio Astrofisico di Torino, Strada Osservatorio 20, Pino Torinese 10025, Italy \inst{1} \\
Laboratory for Astroparticle Physics, University of Nova Gorica, SI-5000 Nova Gorica, Slovenia \inst{2}}

\date{Received / Accepted}
\mail{$\,\,$ S. Mancuso: mancuso@oato.inaf.it, $\,\,$ M.V. Garzelli: garzelli@mi.infn.it .}

\abstract{Faraday rotation measures (RMs) of the polarized emission from extragalactic radio sources occulted by the coronal plasma were used to infer the radial profile of the inner heliospheric magnetic field near solar minimum activity. By inverting LASCO/SOHO polarized brightness ($pB$) data taken during the days of observations on May 1997, we retrieved the electron density distribution along the lines of sight to the sources, thus allowing to disentangle the two plasma properties that contribute to the observed RMs. By comparing the observed RM values to those theoretically predicted by a power-law model of the radial component of the coronal magnetic field, using a best-fitting procedure, we found that the radial component of the inner heliospheric magnetic field can be nicely approximated by a power-law of the form $B_r = 3.76 ~ r^{-2.29} {~ ~ \rm G}$ in a range of heights from about 5 to 14 \rs. Finally, our analysis suggests that the {\it radial} computation of the potential field source surface (PFSS) model from the Wilcox Solar Observatory (WSO), assuming a radial field in the photosphere and a source surface located at $R_{ss} = 2.5$ \rs, is the preferred choice near solar minimum.
}
\titlerunning{Radial profile of the inner heliospheric magnetic field}
\maketitle

\section{Introduction}

Despite its fundamental importance, no direct information is available as yet on the inner heliospheric magnetic field, the innermost {\it in situ} measurements still being those obtained in the seventies by the two Helios probes at $\sim 62$ \rs. Such crucial measurements, in the region where the solar wind is heated, accelerated and decoupled from the coronal plasma, will be only available with the advent of the Solar Probe Plus mission\footnote {Solar Probe Plus website, http://solarprobe.jhuapl.edu/}, planned by NASA to be launched no later than 2018. On the final three orbits, Solar Probe Plus will be able to fly within 8.5 \rs\ from the Sun's surface. Empirical estimates of magnetic fields in the inner heliosphere are only possible very close to the base of the corona (e.g., Lin et al. 2000; Lee 2007), but their outward extrapolation often involve simplistic potential (or force-free) assumptions and the hypothesis of low plasma $\beta$, which may not be appropriate in the outer corona (Gary 2001). Under appropriate assumptions, magnetic field strength estimates in the outer corona have been obtained from the analysis of the radiation emitted during the passage of coronal shock waves (e.g., Mancuso et al. 2003; Vr{\v s}nak et al. 2004; Cho et al. 2007; Bemporad \& Mancuso 2010, Gopalswamy \& Yashiro 2011; Kim et al. 2012). However, one of the best observational methods for obtaining information on the strength and radial profile of the magnetic field in the inner heliosphere still remains the analysis of Faraday rotation measurements of linearly polarized radio signals from galactic or extragalactic radio sources (e.g., Sakurai \& Spangler 1994; Mancuso \& Spangler 1999, 2000; Spangler 2005; Ingleby et al. 2007; Mancuso \& Garzelli 2007; Ord et al. 2007) or from the transmitter of a spacecraft (e.g., Stelzried et al. 1970; P\"atzold et al. 1987; Jensen et al. 2005). For a fairly recent comprehensive review of coronal Faraday rotation observations, see Bird (2007).

In order to understand the physics behind the Faraday effect, it is useful to consider a linearly polarized wave as two counter-rotating circularly-polarized waves. Faraday rotation occurs because the phase velocity of the left circularly polarized component travels faster along the magnetic field than the right circularly polarized component, thus resulting in a net rotation $\Delta \xi $ of the wave's polarization position angle given (in radians) by: 
\begin{equation}
\Delta \xi =\lambda^2 {e^3 \over{2\pi{m_e}^2c^4}}\int_{LOS}{n_e}{\bf B}\cdot{\bf ds} .
\end{equation}
In the above equation, expressed in cgs units, $\lambda$ is the wavelength of the radio signal, $n_e$ is the electron density, {\bf B} is the vector magnetic field, {\bf ds} is the vector incremental path defined to be positive towards the observer, $e$ and $m_e$ are the charge and mass of the electron, and $c$ is the speed of light. The above expression can also be written as $\Delta \xi= \lambda^2 RM$, where $RM$ is the rotation measure (RM), defined to be positive for {\bf B} pointed towards the observer. Essentially, RM yields information on the integrated product of the line of sight (LOS) component of the magnetic field and the electron density. Since Faraday rotation is an integrated quantity and proportional to the product of two independent quantities, in order for the magnetic field component of the observed RM to be determined, there must be an independent determination of the electron density, whether from observations or models. For a spherically symmetric corona, that is, with a perfectly symmetric electron density and magnetic field distributions, the obvious result would be $RM \approx 0$ at all latitudes. This is due to the fact that the LOS component of the magnetic field would reverse at the point of closest approach to the Sun. Overall, the analysis of the Faraday rotation observations reduced by \ms\  showed values different from zero, reaching a maximum value (in absolute magnitude) of $-10.6$ \rmq\ for one of the radio sources at $\sim 6$ \rs. According to the study of \ms\ , this effect was mostly attributed to a non-symmetric distribution of the electron density along the LOS, thus working as a "weighting function" for an otherwise symmetric magnetic field distribution. Although the electron density distribution used in \ms\ was not symmetric, it was still an analytical approximation inspired by the work of Guhathakurta et al. (1996), with two different analytical expressions for the streamer and coronal hole components. In fact, because of this assumption, no attempt was made to actually derive, by a best-fitting procedure, an analytical expression for the radial component of the magnetic field. Moreover, in that work, the position of the magnetic neutral line, which coincides with the heliospheric current sheet, was deduced from a potential field expansion of the photospheric magnetic data from the Wilcox Solar Observatory (WSO) and using a potential field source surface (PFSS) model (Altschuler \& Newkirk 1969; Schatten et al. 1969) with a source surface at a fixed height of 2.5 \rs. Although \ms\  could obtain a fair agreement between RM model and observations, their analysis was only able to support the validity of the P\"atzold et al. (1987) coronal magnetic field model, which was used in input as an approximate expression for the radial magnetic field profile, at the heights covered by the RM observations. 

The aim of this work is to improve the previous analysis of \ms\ by introducing direct information on the electron density distribution along the LOS for the above observations, thus allowing to obtain an estimate of the strength and radial profile of the inner heliospheric magnetic field. In this work, instead of using an analytical model for the electron density distribution along the LOS, we will obtain this quantity empirically through inversion of polarized brightness ($pB$) measurements obtained from observations of the Large Angle and Spectrometric COronagraph (LASCO; Brueckner et al. 1995) telescope aboard the Solar and Heliospheric Observatory (SOHO; Domingo et al. 1995) spacecraft during the same days of observations of the radio sources. In this way, we will be able to infer, by a best-fitting procedure, the radial profile of the magnetic field of the inner heliosphere that best accounts for the observed RMs. As a by-product of our study, we will also be able to evaluate the most suitable PFSS model among the three available models from WSO.

This paper is organized as follows. In Sect. 2, we present some details of the observations and of the data reduction procedure. In Sect. 3, we introduce the magnetic field model. In Sect. 4, we compare the model with observations and discuss our results. Finally a summary and conclusions are given in Sect. 5.

\begin{figure*}
\centering
\epsfig{figure=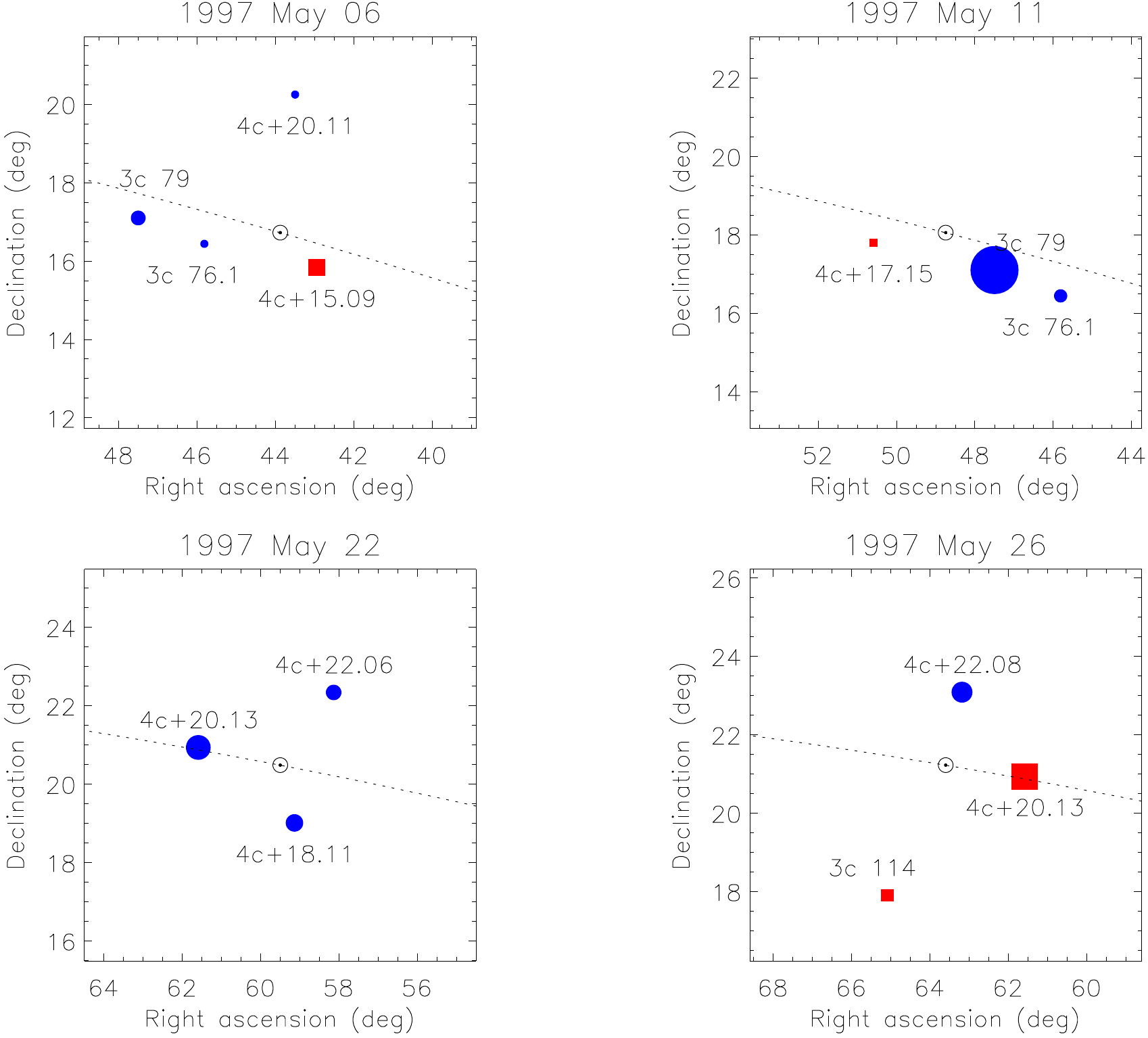,width=13.4cm}
\caption{The above four panels show the location of the radio sources relative to the Sun on each of the four days of observation. The bull's eye symbol indicates the position of the Sun and the dotted line is the ecliptic. The size of the plotted symbol is a rough indicator of the absolute magnitude of the RM. Red squares correspond to positive RMs and blue circles represent negative RMs.}
\end{figure*}

\section{Observations and data reduction}

\subsection{Radio observations}

The observations of the radio sources discussed in this paper were made on four days in 1997 (May 6, 11, 22 and 26) with the Very Large Array (VLA) radio telescope of the National Radio Astronomy Observatory (NRAO). Each source was reobserved far from the Sun at the same frequencies to obtain the intrinsic RM and that due to the interstellar medium but without a coronal contribution. The frequencies used were 1465 MHz with a bandwidth of 50 MHz and 1665 MHz with a bandwidth of 25 MHz and were chosen to have sufficient separation in $\lambda^2$ to allow for an accurate determination of the RMs. Ionospheric calibration was carried out to correct for ionospheric Faraday rotation. Two major factors limiting the accuracy of VLA polarization measurements are the removal of ionospheric Faraday rotation and the correction for instrumental polarization. The task FARAD of the NRAO Astronomical Image Processing System (AIPS) was used for ionospheric RM estimation and correction. This task calculates the appropriate Faraday rotation correction using a phenomenological model of the ionospheric electron density. This model (Chiu 1975), takes as input the monthly average Zurich sunspot number to estimate the ionospheric electron column density. The adequacy of the ionospheric correction was tested by observing four sources far from the Sun, where coronal Faraday rotation is negligible. 
For these sources the standard deviation of the rotation measure fluctuations around the mean was about 
0.3 \rmq\ (see \ms\  for details). This quantity can be considered as a rough estimate of the 
measurement errors introduced by residual ionospheric Faraday rotation or instrumental polarization errors. In both coronal and reference observations, calibration of the instrumental polarization (and subsequent correction) was achieved by observing a calibrator source over a large range in parallactic angle. The procedure used is described in Mancuso \& Spangler (1999). The observations of the radio sources discussed in this paper sampled solar elongations from $\sim 5$ to 14 \rs\ at various heliographic latitudes through 13 different lines of sight (see Fig. 1), with values for the RMs ranging from $-10.6$ to $+3.3$ \rmq. More specific information on the VLA observations used for this work and on the data reduction can be found in \ms\  and will not be repeated here. 

\subsection{White-light observations}

Observations of the extended corona are primarily obtained by white-light coronagraphs that are able to observe coronal structures which are highlighted by the photospheric radiation Thomson-scattered by electrons in the ionized coronal plasma. The observable polarized brightness ($pB$) is defined as the difference in intensity of radiation polarized tangential to the limb of the Sun and radiation polarized normal to the limb. This quantity is directly related to the coronal electron density by 
\begin{equation}
pB(x,y) = \int_{LOS} T(r)n_e(r,\theta,\phi)ds , 
\end{equation}
where $n_e$ is the electron density, $T(r)$ is the Thomson scattering function, and the integral is carried out along the LOS. In general, it is difficult to infer the density distribution in the corona and inner heliosphere from remote sensing techniques since the signals usually derive from the contribution of a superposition of different structures integrated along the LOS in the optically thin intervening plasma. Coronal mass ejections can also drastically affect the $pB$ data as well as disturb the overall structure of the corona in a temporal scale of a few hours. Notwithstanding the above observational caveats, transient phenomena are much less common during solar minimum conditions. Moreover, the overall structure of the corona is observed to be axially symmetric by a high degree during the periods with lower solar activity due to the dominance of the dipolar component of the global magnetic field of the Sun. The electron density distribution can be thus estimated with a good degree of confidence from the observed $pB$ using a technique developed by Van de Hulst (1950) that is particularly appropriate in the near-symmetric density distribution observed during solar minima conditions (e.g., Gibson et al. 1999). In this work, the inversion of the $pB$ data, recorded daily around the Sun by the LASCO C2 telescope that observes the solar corona between about 2 and 6 \rs, was done by fitting the observed radial profiles at steps of $3^\circ$ of heliographic latitude. These $pB$ radial profiles were then inverted with the above mentioned technique to yield electron density radial profiles below about 6 \rs. Finally, radial power fits of the form $n_e(r) = Ar^{-\alpha} + Br^{-2}$ were performed in order to extend the profiles to 1 AU. Once the electron density maps were produced, the data were placed in a rectangular grid with observation times converted to Carrington longitudes and the Carrington maps were finally resampled by using a moving smooth filter and then a moving average, both applied to all pixels. Apart from the assumption of the corona being essentially stationary during the period corresponding to our observations, the above reconstruction of the global coronal electron density implicitly assumes that the observed Thomson scattered radiation is dominated by a relatively thin layer of scattering electrons centered on the point of closest approach to the Sun. This is clearly a zero-order approximation, since the actual $pB$ should be considered as a weighted sum of contributions to the Thomson scattered radiation from electrons along all the line-of-sight. However, it is a fairly safe assumption given that the electron density distribution in the corona falls off with distance as a high-exponent power-law.

We finally mention that although there was also availability of LASCO C3 measurements, probing the corona up to about 30 \rs, during the time interval under examination, these data were not used. In fact, apart from being noisier than the LASCO C2 data, it is known that the contribution from the F (dust) corona cannot be neglected above $\sim $5 \rs\ (e.g., Koutchmy \& Lamy 1985; Mann 1992; Morrill et al. 2006) and the separation between the K and F coronae is not straightforward, thus making the $pB$-based inversion method more problematic and the determination of the electron density more difficult. Finally, in the range covered by the LASCO C3 field of view (about 4 to 32 \rs), it is statistically more probable, with respect to the observations obtained within the LASCO C2 field of view, to find coronal mass ejection (CME) signatures which would significantly distort the observed $pB$ profiles at all heliolatitudes.

\section{Magnetic field model}

The coronal magnetic field during solar minimum activity is dominated by the low-order magnetic multipoles, with the largest contribution coming from the monopole and dipole components. In this work, as in P\"atzold et al. (1987), the radial component of the global magnetic field $B_{rad}(r)$ is thus assumed to have the functional form given by: 
\begin{equation}
B_r ~ = ~ {{B_{01} \over r^3} + {B_{02} \over r^2}} \, .
\end{equation}
The dual power-law form of eq. (3) for $B_r$ is the linear combination of a dipolar field (${\propto r^{-3}}$), representing the dominant component of the global solar magnetic field at solar minimum and a monopolar, solar wind component (${\propto r^{-2}}$) prevailing at large distances from the Sun. Closer to the Sun, higher-order multipole components are present, but their contribution can be ignored at the heights relevant to these RM observations, since such fields fall off rapidly with height. Analysis of Helios data (Mariani et al. 1979) showed that the absolute value of daily averages for the radial magnetic field component scaled as $B_r \propto r^{-2}$ between 0.3 and 1 AU near solar minimum, thus justifying the presence of the monopolar component in eq. (3). Moreover, the radial component of the heliospheric magnetic field, as detected by Ulysses during the same solar minimum as the one investigated in this work, was found to be remarkably constant with latitude, with an average value of $|B_r| \approx 3.1 \times 10^{-5}$ G (Smith \& Balogh, 1995; Balogh et al. 1995). In this paper, the value of $B_{02}$, representing the scale factor for the solar wind component ($\propto r^{-2}$) at great distances from the Sun, was thus fixed by this observational constraint, that is, $B_{02} \approx 1.43$ G $\times$ \rs$^2$ for $B_r $ expressed in Gauss units and $r$ in solar radii (\rs). 

Helios observations also revealed that during solar minimum the slow solar wind of the streamer belt is restricted to a region of about $\pm 20^\circ$ around the heliospheric current sheet. A similar result has been obtained by the Ulysses probe (e.g., Wock et al. 1997) that detected a sharp transition in latitude from slow to fast solar wind. Thus, in the following, we will assume for the sake of simplicity that the slow wind (streamer belt) region is confined in a region within $20^\circ$ above and below the heliospheric neutral line, with the fast wind occurring beyond this latitudinally bounded region. Another important parameter is the Alfv\'en radius $r_A$, that is, the distance at which the solar wind becomes super-Alfv\'enic in the outer corona. Below $r_A$, the field is strong enough to control the plasma flow and cause the solar wind to corotate with the solar wind that has no azimuthal component in a corotating frame. Above $r_A$, the plasma is released from corotation and the field wraps up following, approximately, an Archimedean spiral (Parker 1958). For $r \leq r_A$, in a spherical coordinate system ($r,\theta,\phi$) the azimuthal component of the magnetic field, $B_{\phi}$, is null. Beyond $r_A$, $B_\phi$ is different from zero, slowly growing with $r$ as 
$B_\phi \propto \Omega_{\odot} (r- r_A) \cos\theta/u_{sw} $,
thus being a function of the coronal rotation rate $\Omega_{\odot}$, the solar wind speed $u_{sw}$, and of heliolatitude $\theta$. The value of $r_A$ remains uncertain as yet, but it has been estimated as ranging from $\sim10$ \rs\ over the poles to $\sim 30$ \rs\ over the equator (Scherer et al. 2001). In the following, $r_A$ will be modeled by a step function assuming a value of $r_A = 30$ \rs\ in the slow wind region around the equator and $r_A = 10$ \rs\ in the fast wind region towards the poles. 

Observations of the solar coronal rotation rate in solar cycle 23 have shown that the extended corona is more rigid than the photosphere (e.g., Giordano \& Mancuso 2008; Mancuso \& Giordano 2011, 2012) but still sensibly latitude-dependent. In this work, the coronal rotation rate has been approximated by the functional form given by Chandra et al. (2010) for the soft X-ray corona in 1997 from images obtained with the soft X-ray telescope (SXT) on board the $Yohkoh$ solar observatory. As for the solar wind speed, the value for the slow component was chosen on the basis of the average speed of the slow wind at 1 AU as deduced from the data obtained by the Solar Wind Experiment (SWE; Ogilvie et al. 1995) on the Wind spacecraft during May 1997. As for the fast solar wind at 1 AU, we relied on the analysis of Quemerais et al (2007), who deduced the velocity profiles for Spring 1997 from the combination of SWAN, LASCO, and IPS data. We remark that although we took into account, for completeness, the contribution of the azimuthal component of the magnetic field to the observed RM, this amount remains practically negligible (of the order of a few percent) in the range of heights considered in this work. 

Finally, a necessary ingredient in coronal RM studies is the knowledge of the location of the heliospheric current sheet with respect to the LOS to the occulted radio sources. This information is important to infer the polarity of the magnetic field along the integration path to the different sources, especially near the solar equator. This is customarily done through the use of PFSS models. Because of their simplicity, PFSS models are still used extensively within the solar and heliospheric communities. Riley et al. (2006) demonstrated that PFSS models are especially applicable during solar minimum activity and that can generate solutions that closely match those generated by MHD models for cases when time-dependent phenomena are negligible.  PFSS models assume that the magnetic field can be described as the negative gradient of the scalar potential $\Phi$ which satisfies Laplace's equation $\nabla^2\Phi = 0$, thus implying that $\nabla\times {\bf B} = 4\pi{\bf J}/c = 0$, so that current densities {\bf J} are neglected. Laplace's equation for $\Phi$ is solved in the space between the photosphere and an outer spherical shell (the so-called {\it source surface}, with radius $R_{ss}$), yielding a solution in terms of Legendre polynomials. The so-obtained scalar potential is uniquely determined given inner and outer boundary conditions at the photosphere and at the source surface, respectively. The magnetic neutral line, when propagated outwards, can be used as a proxy for the latitudinal location of the warped heliospheric current sheet, and this location can be readily and reliably obtained through the PFSS models. The coronal magnetic field calculated from photospheric field observations with the PFSS models is tabulated by the WSO and available online at their website with two different extrapolation methods. The {\it classic} computation, which locates the source surface at 2.5 \rs, assumes that the photospheric field has a meridional component and requires a somewhat {\it ad hoc} polar field correction in order to better match the {\it in situ} observations at 1 AU. The {\it radial} computation, with $R_{ss}$ located at 2.5 \rs\ and 3.25 \rs, assumes that the field in the photosphere is radial and requires no polar field correction. The position of the magnetic neutral line, separating regions of opposite polarity, is simply given by the contour of zero magnetic field strength. In our model, the magnetic field is thus assumed to be almost purely radial above the source surface and a function only of radius except for the polarity switching sign across the neutral line. The location of the heliospheric neutral sheet and, consequently, the polarity of the magnetic field, will be inferred from the position of the magnetic neutral line as deduced from the three available PFSS models from WSO.

\section{Results and discussion}

In the following, we will implement three sets of models. In the first set of models (Case A), in order to reproduce the observed RMs, we will use the functional form for the radial component of the heliospheric magnetic field given by eq. (3) with $B_{01}$ left as a free parameter. We will further assume that the radial variation of the electron density $n_e$ with heliographic coordinates can be reproduced with accuracy by the $pB$ inversion method outlined in Sect 2.2. This hypothesis might be questionable in that it would require both perfect calibration of the available LASCO C2 data and that the extrapolation to higher distances is accurate enough. In a second set of models (Case B), we will set $B_{01} = 0$ and investigate the possibility that the electron density $n_e$ has been obtained only up to an unknown multiplicative factor $C_n$, so that its magnitude needs some adjustment (that is, $n_e \rightarrow C_n n_e$) to be in agreement with the actual RM observations (see also the discussion in Ingleby et al. 2007). By setting the parameter $B_{01}$ to zero and letting $C_n$ to be the only free parameter, we will test the possibility that the radial magnetic field, down to a few solar radii from the Sun, might be simply described by a solar wind component (${\propto r^{-2}}$) (e.g., Vr{\v s}nak et al. 2004; Spangler 2005; Ingleby et al. 2007). Finally, in the most general case (Case C), both $C_n$ and $B_{01}$ will be left as free parameters. All these models will be separately tested by assuming the three available PFSS computations from WSO (i.e., the {\it classic} computation with $R_{ss} = 2.5$ \rs\ and the {\it radial} computation with $R_{ss} = 2.5$ \rs\ or 3.25 \rs). 

The unknown parameters in the above described models will be evaluated by minimizing the weighted sum-of-squared-residuals:
\begin{equation}
\chi^2 = {\sum_i{\omega_i(RM_{obs,i}-RM_i)^2}}, 
\end{equation}
where $RM_{obs,i}$ is the observed RM of the $i$-th LOS, $RM_i$ is the model RM for the same LOS, and \omi\ is a weight assigned to each measurement depending on its uncertainty. Actually, the proper choice of \omi\ is a critical issue in our statistical analysis. In fact, the available uncertainties \srmi\ $ =$ \srmo\ (of order $\sim 0.1-0.3$ \rmq\ for most observations) for $RM_{obs,i}$ as quoted in \ms\ , were assumed to be solely due to radiometer noise. The RM fluctuations attributable to coronal turbulence and MHD waves, \drm, can be, however, much higher than \srmi. For example, RM measurements from the Helios data showed RM fluctuations up to \drm\ $ \sim 2$ \rmq\ with a time scale of roughly half an hour (Hollweg et al. 1982; Bird et al. 1985). Similarly, Sakurai \& Spangler (1994) set an upper limit of \drm\ $ \sim 1.6$ \rmq\ for a radio source observed during solar maximum whose LOS passed within $\sim 9$ \rs\ of the Sun, although Mancuso \& Spangler (1999), in an analogous analysis near solar minimum, found \drm\ $ \lesssim 0.4$ \rmq.  The overall level of turbulent RM fluctuations is thus not well constrained as yet and might be dependent as well on the particular phase of the solar cycle. Since there was no simultaneous independent estimate of the contribution of RM fluctuations attributable to coronal turbulence and MHD waves during the days of observations, the uncertainties \srmi as quoted in \ms\ , may underestimate by some factor the {\it true} uncertainties in each measurement, given by \srmi$^* = (\sigma_{RM,i}^2 + \delta RM^2)^{1/2}$. As a consequence, the true uncertainties \srmi$^*$ remain actually unknown. 
According to the previous discussion, in the process of \chit\ minimization for the evaluation of the unknown parameters, we have considered different \omi\ estimates, by proceeding through three different steps, as explained below. 

First of all, in Step 1, we analyzed the RM data by considering as their total uncertainties \srmi\ the ones that can be ascribed to radiometric noise only, i.e., we completely neglected the possible contribution of coronal turbulence and MHD waves to RM fluctuations. We minimized the $\chi^2$ for each of the cases A, B and C, and for each available PFSS model. We then computed the reduced chi-squared $\chi^2_\nu \equiv \chi^2/\nu$ by taking into account the number of degrees of freedom ($\nu$) in the different cases (12 for Cases A and B, and 11 for Case C, due to the different number of estimated parameters). We also estimated the goodness of fit by computing the probability of observing a minimum of the distribution that arises from eq. (4) by varying the model parameters, larger than the one actually observed, assuming that this minimum is actually distributed according to a $\chi^2$ with $\nu$ degrees of freedom (e.g., Garzelli \& Giunti 2002). Calling $\alpha$ this probability, that in practice represents the area of the right tail of the $\chi^2$ distribution in the interval between [$\chi^2_{min}$ and $+\infty$), we say that the fit is acceptable at 100\% $\alpha$ confidence level. In all cases, the reduced chi-squared for the best-fit points were $\chi^2_\nu > 30$, thus yielding a very poor $\alpha$. These results clearly suggest that the uncertainty component due to radiometric noise largely underestimates the total uncertainty, at least for some of the sources, and that the uncertainties \drm\ due to RM fluctuations cannot be neglected in our study.
Secondly, in Step 2, in the absence of an experimental estimate of \drm\ for the measurements at hand, we considered the range of possible \drm\ estimates provided in literature (see the discussion above), in order to obtain a gross estimate of the typical order of magnitude of \drm. Taking into account the interval spanned by these estimates, we decided to adopt an average value \drm\ $ = $ 1  \rmq, that we assumed to remain constant during the days of our RM observations. We also assumed that, as expected in a realistic situation, the RM fluctuations do not act by systematically increasing or by systematically decreasing all RM observed values in the same direction, but may act in a different way on each RM observation, i.e., they do not introduce any correlation between the RM observations of different sources. We thus replaced the uncertainties \srmi\ = \srmo\ used in Step 1 with the uncertainties \srmi$^* = (\sigma_{i,obs }^2 + \delta RM^2)^{1/2}$, finding \chitn\ values compatible with 1, thus leading to a more meaningful statistical analysis. In this configuration, we were able to identify the best PFSS model among those proposed by WSO (that turned out to be the same for all three cases A, B and C), the best-fit parameters for each case, and the uncertainties on them at 90\%, 95\% and 99\% confidence levels. These uncertainties are given in the following as confidence intervals in the cases where we just vary a single parameter (cases A and B), and as confidence regions when, instead, we vary more parameters at the same time (Case C). 
Finallly, in Step 3, we considered a statistical analysis with uniform weighting, i.e., with \omi\ arbitrarily set to 1, so to yield an unweighted least-squares. The results of this analysis turned out to be similar to those of the analysis of Step 2, thus proving the marginal role of \srmo\ on the analysis' outcome, with respect to the dominant contribution due to turbulence and MHD waves (i.e, \srmi$^* \approx \delta RM \gg \sigma_{i,obs}$).

\begin{figure}[t!]
\centering
\includegraphics[bb=0 0 503 503,width=12cm]{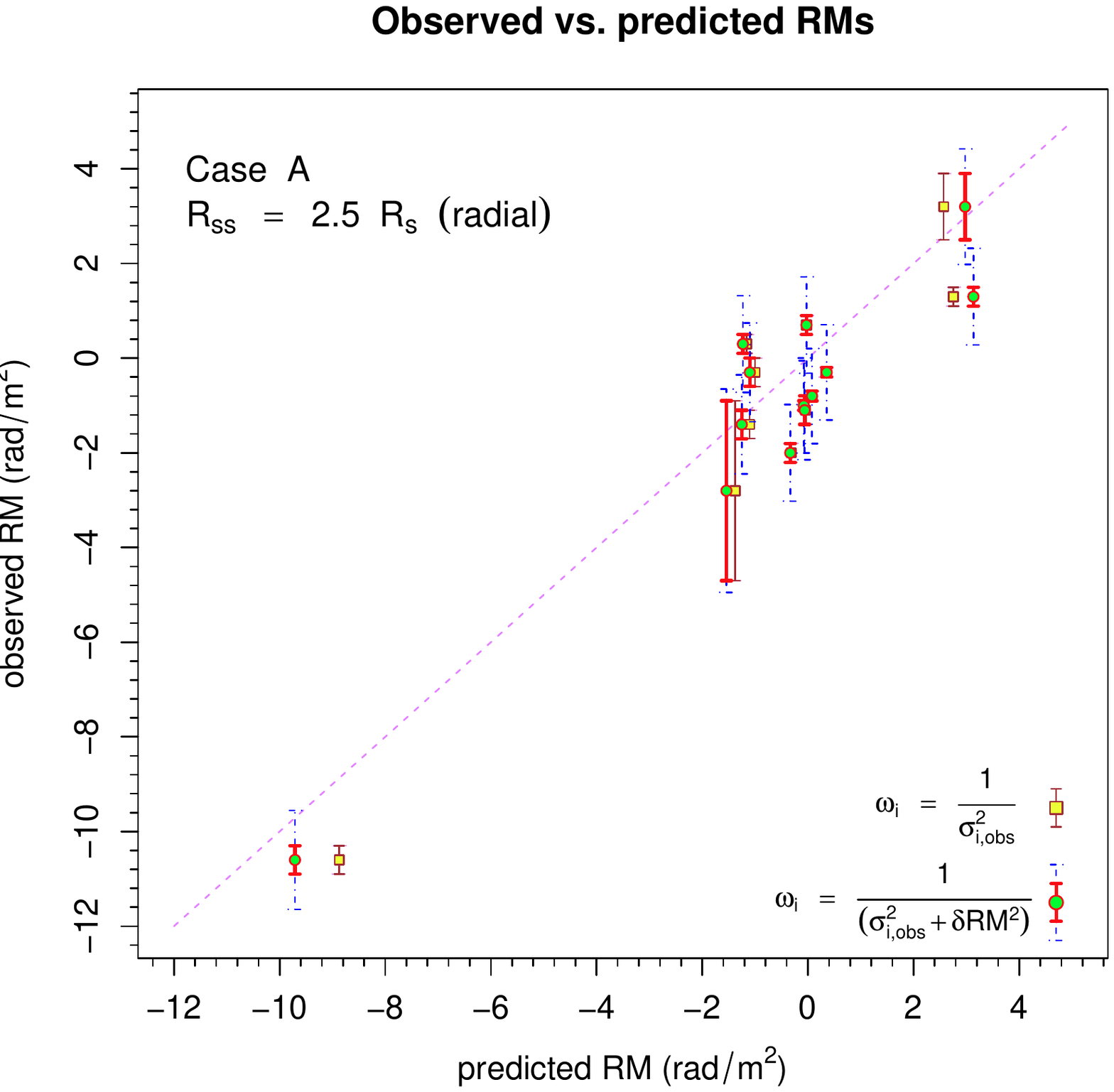}
\caption{Scatter plot of predicted RMs versus observed RMs in the best-fit parameter configuration, using a model with $B_{01}$ left as free parameter and setting $C_n = 1$. Results are displayed for both cases of non-uniform weights, after the two analyses described in Step 1 (yellow squares) and Step 2 (green circles), respectively. The dashed 1:1 line shows the diagonal, corresponding to the ideal situation where predicted and measured values are equal. The solid (red and brown) error bars denote the \srmi\ uncertainties for each measurement due to radiometer noise as quoted in Mancuso \& Spangler (2000), whereas the dashed (blue) errorbars denoted the \srmi$^*$ total uncertainties obtained after adding the contribution of RM fluctuations, estimated to be  \drm\ $ = $ 1 \rmq\ for the sake of this analysis.}
\end{figure}

In our opinion, the results obtained using the weighting criteria of Steps 2 and 3 are actually, from a physical point of view, more representative of the real situation, insofar the $true$ uncertainties are most probably dominated by the contribution of RM fluctuations attributable to coronal turbulence and MHD waves. In the following, we will therefore concentrate our discussion mostly on the results concerning Steps 2 and 3.
We remark, however, that the above conclusion is strictly valid only in the assumption (yet to be verified in the range of heliocentric distances considered in this work) of  \drm $ \sim $ constant both in latitude and distance from the Sun.

\begin{figure}[t!]
\centering
\includegraphics[bb=0 0 503 503,width=12cm]{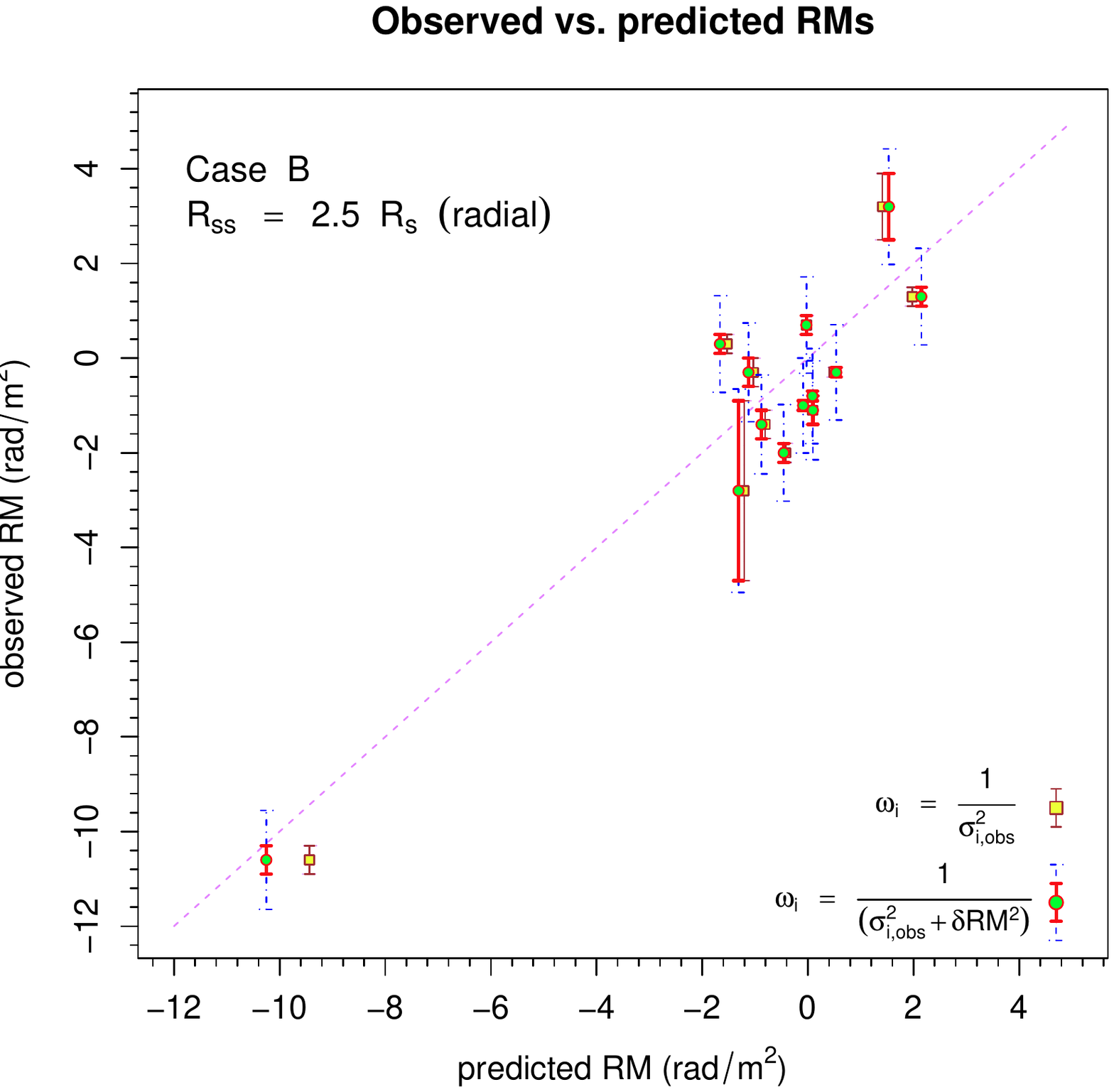}
\caption{\label{scatter2} Same as Fig. 2, but using a model with $C_n$ left as free parameters and setting $B_{01}=0$. }
\end{figure}

\begin{figure}[t!]
\centering
\includegraphics[bb=0 0 503 503,width=12cm]{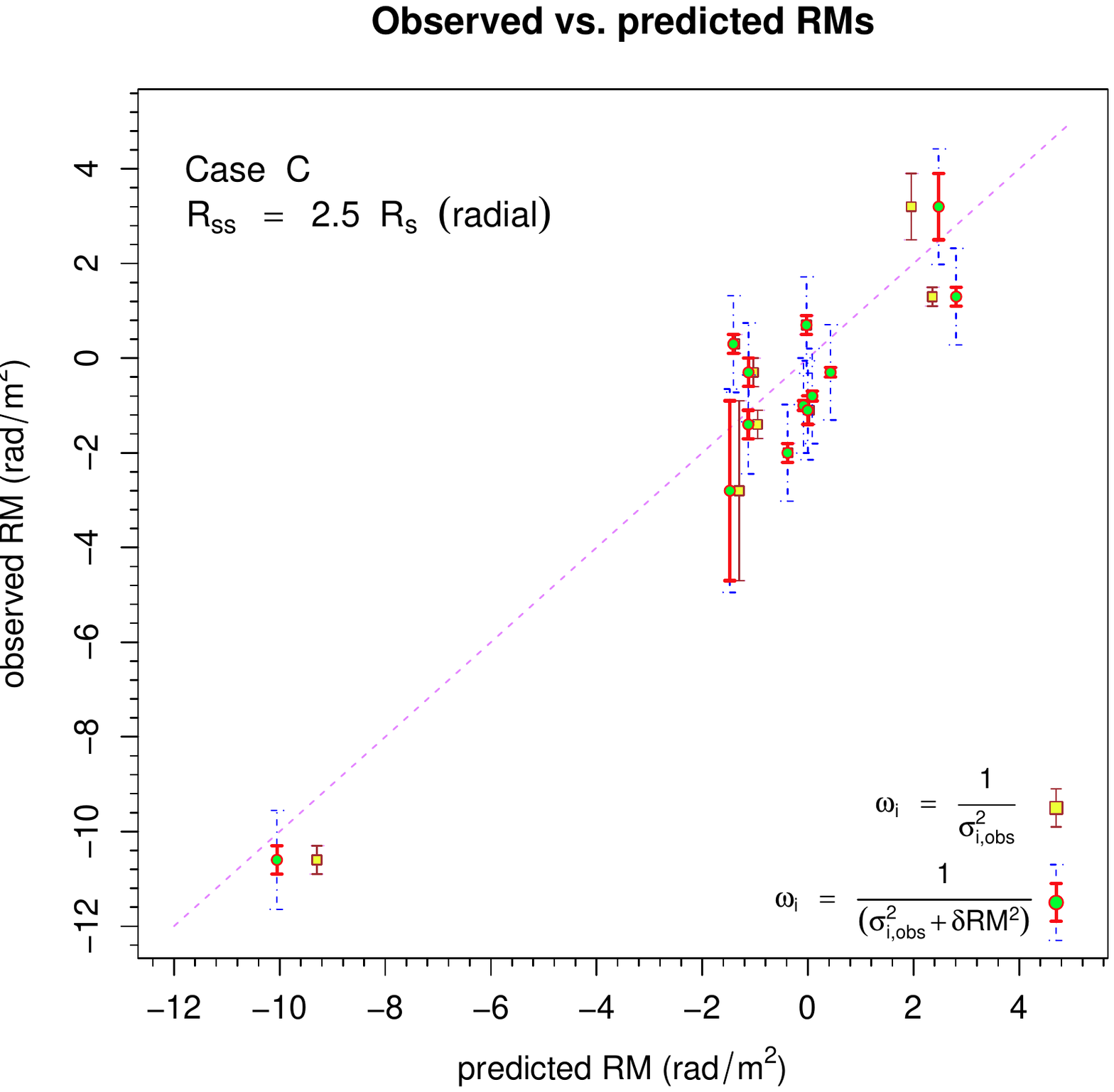}
\caption{\label{scatter3} Same as Fig. 3, but using a model with both $B_{01}$ and $C_n$ left as free parameters. }
\end{figure}

\subsection{Model with $B_{01} \neq 0$ and $C_n = 1$ (Case A)}

To find the regions of parameter space consistent with the RM observations, we performed a first grid search allowing the two unknown parameters $B_{01}$  and $R_{ss}$ to vary within a plausible range and fixing the multiplicative factor $C_n$ to unity (that is, assuming that the radial variation of the electron density with heliographic coordinates is reproduced with accuracy). 
The ranges of the model parameters that define our search space were constrained to be within a plausible interval of $0 < B_{01} < 20$ (with $B_{01}$ expressed hereafter in units of G $\times$ \rs$^3$) at Steps of 0.01 \rs.  
The result of the best-fitting procedure yielded a very well defined $\chi^2$ minimum, thus allowing an accurate determination of the best-fit parameters. 
The best-fit parameters to the observed data for Case A 
are $B_{01} = 10.54$ and $B_{01} = 10.95$ for the two weighing criteria explained above (Steps 2 and 3) and for the {\it radial} computation with $R_{ss} = 2.5$ \rs. The {\it radial} computation with $R_{ss} = 3.25$ \rs\ yields higher $\chi^2$ values, while the {\it classic} computation with $R_{ss} = 2.5$ \rs\ yields a very high  $\chi^2$ value, hinting to $B_{01} = 0$.
How well a model performs compared to the measurements is conveniently assessed from the scatter plot of observations versus predictions. Tighter scatter of the data points around the 1:1 line obviously indicates more accurate model fit. In Fig. 2, we show a scatter plot of predicted RMs versus observed RMs for Case A obtained by using the best-fitting parameters and the two non-uniform weighting criteria listed in Steps 1 and 2. The estimated RM values look to be nicely distributed around the line 1:1 of perfect fit for all the range of measured RM values. The largest discrepancies between the results of the models of Steps 1 and 2, that differ because of the RM uncertainties attributed to each observation, also shown in the plot, are visible in the RM predictions for those sources that have a larger absolute experimental RM value. In particular, adopting the model of Step 2 appears to greatly improve the agreement between theory and experiment for the source with an highly negative measured RM ($<$ -10 \rmq).

\subsection{Model with $B_{01} = 0$ and $C_n \neq 1$ (Case B)}

In the model presented in the previous section (Case A), we assumed that the radial variation of the electron density with heliographic coordinates could be reproduced with accuracy by the $pB$ inversion method applied to the white-light data, thus requiring perfect calibration of the available data and accuracy in the extrapolation of the LASCO C2 data to higher distances. As already mentioned, however, the electron density $n_e$ can be probably retrieved, at best, only up to an unknown multiplicative factor $C_n$, representing an adjustment factor introduced to produce agreement with the actual RM observations (see also the discussion in Ingleby et al. 2007). 
Accordingly, in the second model (Case B), we performed a grid search by setting $B_{01} = 0$ (that is, assuming no dipolar component), allowing 
the parameter $C_n$ to vary within a plausible interval of $0.3 < C_n < 3$ at Steps of  0.01 \rs. 
This model was implemented to test the possibility that the radial component of the magnetic field might be already ${\propto r^{-2}}$ at distances as low as a few solar radii from the Sun (e.g., Vr{\v s}nak et al. 2004; Spangler 2005; Ingleby et al. 2007 ). 
The best-fit parameter to the observed data for Case B is $C_n = 2.00$ for the {\it radial} computation with $R_{ss} = 2.5$ \rs. Again, the {\it radial} computation with $R_{ss} = 3.25$ \rs\ yields higher  $\chi^2$ values, while the {\it classic} computation with $R_{ss} = 2.5$ \rs\ yields a very high  $\chi^2$ value, hinting to low $C_n$ values. 
The scatter plot of predicted versus observed RMs, with predictions obtained by using the best-fitting parameters for this case, assuming a PFSS radial computation with $R_{ss} = 2.5$ \rs\ is shown in Fig. 3. Considerations similar to those  concerning Fig.~2 are valid (see the discussion at the end of the previous subsection).

\subsection{Model with $B_{01} \neq 0$ and $C_n \neq 1$ (Case C)}

In the last model (Case C), we further performed a grid search allowing 
the two unknown parameters $B_{01}$ and  $C_n$, to vary within the interval of $0 < B_{01} < 40$, and $0.3 < C_n < 3$. Also in this case, a minimum $\chi^2$ was very well defined allowing us to accurately determine the best-fit parameters. The model parameters that provide the best fit to the observed data 
are $B_{01} = 4.84$ and $C_n = 1.39$ for $\omega_i = 1/\sigma_{RM,i}^{* \,\,2}$, and $B_{01} = 6.46$ and $C_n = 1.28$  for $\omega_i = 1$. Again, the best-fit was obtained for the {\it radial} computation with $R_{ss} = 2.5$ \rs.  
In Fig. 4, we show the scatter plot of predicted versus observed RMs obtained by using the best-fitting parameters. 
Considerations similar to those concerning Fig.~2 apply even in this case.

\subsection{Discussion and comparison with other models}

  The model results are summarized in Table 1.
 For comparative purposes, for each considered model, the parameters $C_n$ and $B_{01}$ that provided the best-fit to the observed RM's are tabulated together with the corresponding best-fit \chitn, for both cases of non-uniform (\omi\ $ = 1/\sigma_{RM,i}^{2}$ and \omi\ $ = 1/\sigma_{RM,i}^{*\,\,2}$) and uniform (\omi\ $\equiv 1$) weighting, corresponding to the three steps outlined above for the evaluation of the uncertainties. We also report the estimate of $\alpha$, for those models for which it is larger than 1\%. We remark that the values of this variable strictly depend on the above hypotheses concerning the RM uncertainties (see the discussion before subsection 4.1), thus it has to be considered with some care. For the model on the last line of Table 1 (classic computation with $R_{ss} = 2.5$), we found that the $\chi^2$ is minimized by choosing very high values of the $B_{01}$ component, beyond the physical limits expected to be plausible on the basis of present knowledge, so this model has probably to be ruled out.

\begin{table}[t!]
\caption{The best-fit models. See text for more detail.} 
\centering\tabcolsep=1.2mm
\begin{tabular}{@{}ccccccccccccccccc@{}} 
\hline\hline
&& \multicolumn{2}{c}{} &
\multicolumn{4}{c}{$\omega_i = 1/\sigma_{RM,i}^{2}$ (Step 1)} & \multicolumn{4}{c}{$\omega_i = 1/\sigma_{RM,i}^{* \,\,2}$ (Step 2)} & \multicolumn{5}{c}{$\omega_i \equiv 1 $ (Step 3)}\\ 
\cmidrule{1-1} \cmidrule{3-3} \cmidrule{5-7} \cmidrule{9-12} \cmidrule{14-17} 
Case && PFSS model && $B_{01}$ & $C_n$ & $\chi^2$/$\nu$ && $B_{01}$ & $C_n$ & $\chi^2/\nu$ & $\alpha$ && $B_{01}$ & $C_n$ & $\chi^2$/$\nu$ & $\alpha$ \\
\cmidrule{1-1} \cmidrule{3-3} \cmidrule{5-7} \cmidrule{9-12} \cmidrule{14-17} 

A && radial 2.50  && 8.61 & 1.00 & 446.6/12   && 10.54 &1.00 &13.39/12 & 34.2\% && 10.95 & 1.00 & 15.17/12 &23.2\% \\ 
A && radial 3.25  && 6.77 & 1.00 & 547.8/12   && 9.64  &1.00 &18.45/12 & 10.2\% && 10.15 & 1.00 & 20.59/12 &5.7\% \\ 
A && classic 2.50 && 5.54 & 1.00 & 1333.2/12  && 0.00  &1.00 &105.4/12 &        && 0.00 & 1.00 & 261.0/12 & \\ \\

B && radial 2.50  && 0.00 & 1.84 & 443.3/12   && 0.00  &2.00 &14.13/12 & 29.2\% && 0.00 & 2.04 & 17.24/12 &14.1\% \\ 
B && radial 3.25  && 0.00 & 1.63 & 548.7/12   && 0.00  &1.88 &20.19/12 & 6.4\%  && 0.00 & 1.93 & 24.07/12 &2.0\% \\ 
B && classic 2.50 && 0.00 & 0.63 & 1307.8/12  && 0.00  &0.58 &86.73/12 &        && 0.00 & 0.32 & 108.2/12 & \\ \\

C && radial 2.50  && 2.65 & 1.48 & 437.5/11   && 4.84  &1.39 &12.94/11 &29.7\%  && 6.46 & 1.28 & 14.90/11 &18.7\% \\ 
C && radial 3.25  && 2.80 & 1.30 & 545.7/11   && 7.85  &1.10 &18.45/11 &7.3\%   && 11.20 & 0.95 & 20.57/11 &3.8\% \\
C && classic 2.50 && $>$ 45 & 0.30 & $>$ 660/11  && $>$ 45 &0.20 & $>$ 55/11 & && $>$ 45 & 0.10 & $>$ 88/11 & \\ \hline
\end{tabular} 
\end{table}

Overall, among the nine considered models (see Table 1), the three of them obtained by assuming the position of the magnetic neutral line as deduced from the {\it radial} computation with $R_{ss} = 2.5$ \rs\ always yielded the smallest minimum  \chitn\ for all weighting criteria. In case of the two weighting criteria of Steps 2 and 3, we obtained values of  \chitn\ at the minima small enough to allow for a meaningful statistical analysis ($\alpha >$  1 $-$ 2 \%). The models computed by assuming the {\it radial} computation with $R_{ss} = 3.25$ \rs\ also yielded well-defined $\chi^2$ minima within the allowed parameter space, although with larger $\chi^2$ minimum values and with smaller $\alpha$ values with respect to the {\it radial} computation with $R_{ss} = 2.5$ \rs. Finally, the {\it classic} computation always yielded the largest minimum $\chi^2$ (often at the border of the parameter space) and the worst $\alpha$'s among the three available WSO models. These results allow us to clearly draw our first and more robust conclusion, that is, that the {\it radial} computation of the PFSS model from WSO with a source surface located at 2.5 \rs\ is indeed the preferred choice, at least near solar minimum. 

Looking back at Table~1, it turns out that the minimum \chit\ for Case C is always smaller than that for Case A and B for all weighting criteria. We remark, however, that this inequality is no longer true when comparing Case C and Case A by considering the \chitn, instead of the \chit, due to the different $\nu$, although in both cases \chitn\ $\sim 1$ for the {\it radial} computation with $R_{ss} = 2.5$ \rs. Related to these facts, the $\alpha$ value for Case A turns out, indeed, to be slightly higher than the one of Case C. However, the difference is quite small for both weighting criteria of Steps 2 and 3, and thus we believe that both fits actually deserve full consideration. On the other hand, the $\alpha$ value of Case B appears slightly worse, especially in the case of the weighting criterium of Step 3. Following the indications of our statistical analysis, we consider, for the sake of comparison of our results with other magnetic field estimates in literature, both the models of Case A and C. On the other hand, taking into account the intrinsic  uncertanties and assumptions related to the $pB$-inversion method used in this work, we are aware that our density estimate certainly needs {\it some} adjustment, at best by means of an enhancement or depletion factor. In fact, we believe that the slightly larger $\alpha$ value for Case A with respect to Case C is just an artifact related to the fact that our $\chi^2$ distributions are shallow. This impression is confirmed by comparison of our results in the two cases with independent work already published in the literature, as explained in the following.

Our best-fit model for Case A, expressed as a power-law fit valid in the range between about 5 and 14 \rs, is 
\begin{equation}
B_r ~ = ~ {{10.54 \over r^3} + {1.43 \over r^2}} \approx ~ 7.40 ~ r^{-2.47} { ~ ~ \rm G},
\end{equation}
whereas our best-fit model for Case C, expressed as a power-law fit valid in the range between about 5 and 14 \rs, is 
\begin{equation}
B_r ~ = ~ {{4.84 \over r^3} + {1.43 \over r^2}} \approx ~ 3.76 ~ r^{-2.29} { ~ ~ \rm G}.
\end{equation}

\begin{figure*}[t]
\centering
\epsfig{figure=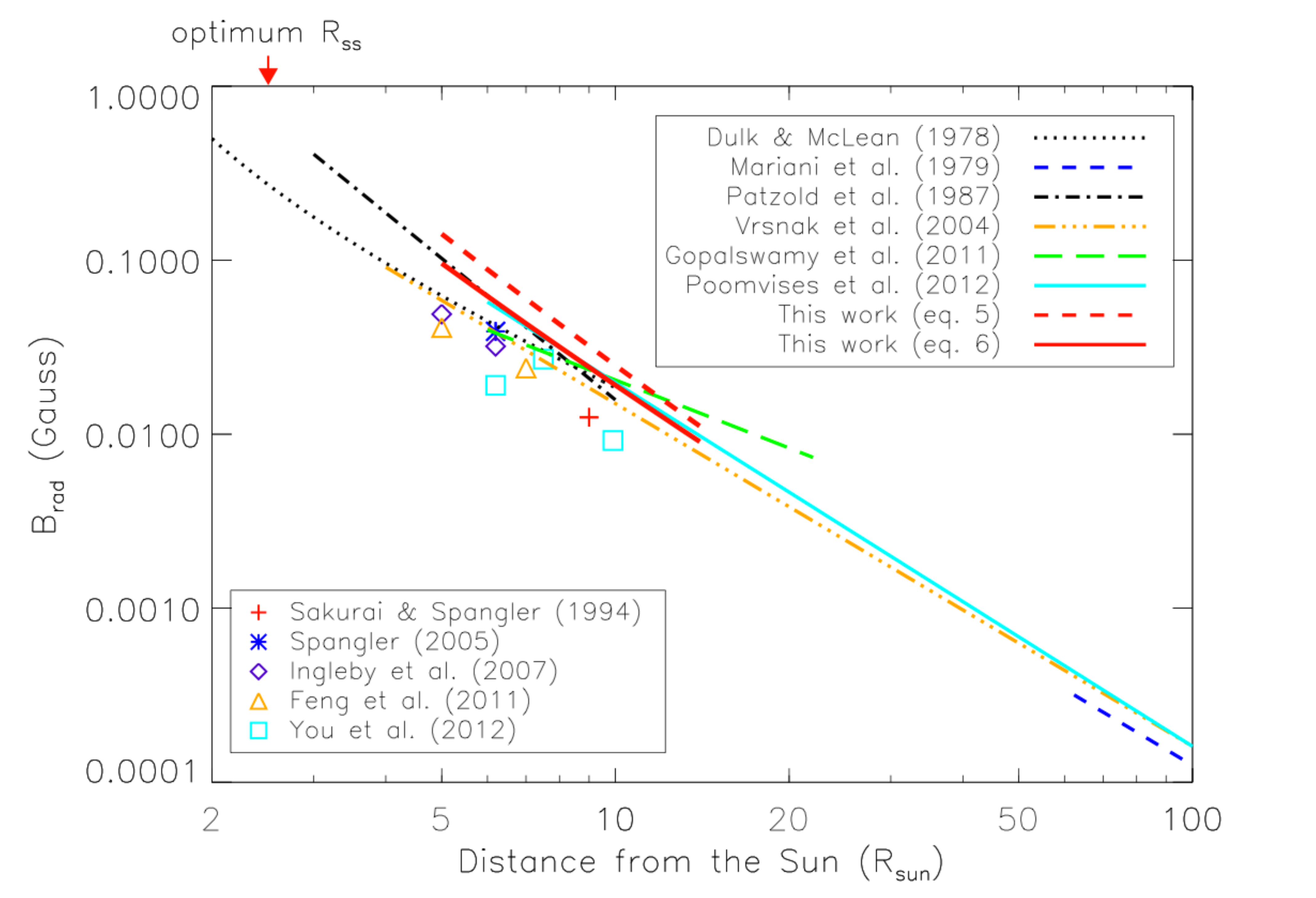,width=14cm}
\caption{Comparison of the result from this work, eqs. (5) and (6), with various estimates (plotted with different symbols and colors) for the radial profile of the magnetic field strength in the inner heliosphere.}
\end{figure*}

Several empirical formulas have been introduced in literature in the past decades to model the radial magnetic field profile. In Fig. 5, we compare the magnetic field radial profile as derived from this work, given by eq. (5) and (6), with the radial profiles obtained in the inner heliosphere by Dulk \& McLean (1978), P\"atzold et al. (1987), Vr{\v s}nak et al. (2004), Gopaswamy \& Yashiro (2011) and Poomvises et al. (2012) with different techniques. Specifically, Dulk \& McLean (1978) derived, from radio observations, a radial profile given by $B_r= 0.5 ~ (r - 1)^{-1.5}$ G for $r < 10$ \rs. The estimate by P\"atzold et al. (1987), $B_r= 7.9 ~ r^{-2.7}$ G, valid in the range between 3 and 10 \rs, was also obtained through Faraday rotation measurements but using data from Helios. Vr{\v s}nak et al. (2004) used information on the band splitting of type II radio bursts, obtaining $B_r = 1.4 ~ r^{-1.97}$ G from the corona up to 1 AU. In the same plot, we also show further estimates of the magnetic field strength in the inner heliosphere with various symbols (Sakurai \& Spangler 1994; Spangler 2005; Ingleby et al. 2007; Feng et al. 2011; You et al. 2012). From Fig. 5, it can be appreciated that the magnitude and variation trend of the radial component of the magnetic field as obtained in our analysis, Case C, is in fair agreement with the other estimates and in particular with the result from P\"atzold et al. (1987), at least in the range of heights around 5 \rs, being somewhat flatter overall. 
Moreover, the match with the estimate of Poomvises et al. (2012), obtained with a completely different method, i.e., from the analysis of the standoff distance of a CME-driven shock observed by the {\it  Solar Terrestrial Relations Observatory} ({\it  STEREO}) spacecraft, is particularly interesting.
On the other hand, the magnetic field profile we obtained instead by assuming $C_n = 1$ (Case A) appears to overestimate all other results, although its extrapolation at higher heights is still in fair agreement with the estimate of Poomvises et al. (2012). 
This substantiates our hypothesis that the $n_e$ as estimated by means of  the $pB$-inversion method needs actually some correction (as suggested by the introduction of the enhancement factor $C_n$), and that our best estimate of the radial magnetic field profile in the inner heliosphere is actually given by eq. (6) and not eq. (5).    
Finally, we remark that the radial profile obtained by Gopalswamy \& Yashiro (2011), who analyzed the shock stand-off distance and the radius of curvature of a flux rope associated to a CME event, is instead  much flatter with respect to our profile. When comparing the above radial magnetic field estimates, however, we remind that all of them were obtained through remote sensing techniques and are therefore strongly influenced by the assumed (or, at best, estimated) electron density distribution along the LOS.

\section {Summary and conclusions}

Faraday rotation measures (RMs) estimated by \ms\  along thirteen LOS to ten extragalactic radio sources occulted by the corona were used to constrain the inner heliospheric magnetic field around solar minimum. Since RM observations basically probe the electron density-weighted magnetic field strength along the integration path, the crucial point in this kind of analysis is obtaining a reliable estimate of the intervening electron density distribution along each LOS. 

By inverting LASCO/SOHO $pB$ data taken during the days of observations, we were able to disentangle the two plasma properties that contribute to the observed RMs, thus allowing the radial component of the inner heliospheric magnetic field, assumed to have a simple analytical form, to be uniquely determined. By comparing observed and model RM values, using a best-fitting procedure, we found that this profile can be nicely approximated by a power-law of the form $B_r = 3.76 ~ r^{-2.29} { ~ ~ \rm G}$, in a range of heights spanning from about 5 to 14 \rs. 
In order to obtain the above estimate, an enhancement factor of  $C_n \approx 1.3$  was required for the electron density $n_e$ as inferred by means of the $pB$-inversion technique.
The magnitude and variation trend of the radial component of the magnetic field as obtained in our analysis is in fair agreement with previous estimates. 
Finally, our analysis suggests that the {\it radial} computation of the PFSS model from WSO with a source surface located at $R_{ss} = 2.5$ \rs\ is the preferred choice near solar minimum.

Future direct measurements of the magnetic field from the magnetometers on board NASA's Solar Probe Plus mission are expected to further constrain the magnetic field radial profile in the range of heights investigated in this work.

\begin{acknowledgements}
The authors would like to thank the anonymous referee for the very helpful comments and suggestions that  significantly improved the paper.
M.V.G. is grateful to C. Giunti for many useful discussions and suggestions concerning the statistical analysis of the data shown in this paper. The SOHO/LASCO $pB$ data used in this work are produced by a consortium of the Naval Research Laboratory (USA), Max-Planck-Institut f\"ur Aeronomie (Germany), Laboratoire d'Astronomie Spatiale (France), and the University of Birmingham (UK). SOHO is a project of international cooperation between ESA and NASA. The NRAO is a facility of the NSF operated under cooperative agreement by Associated Universities, Inc. Wilcox Solar Observatory data used in this work were obtained via the WSO website, courtesy of J. T. Hoeksema. The WSO is supported by NASA, the National Science Foundation, and ONR. 
\end{acknowledgements}

\end{document}